\documentclass[aps,prl,twocolumn,groupedaddress]{revtex4}

\usepackage{graphicx}

\begin{document}

\title{Weak- to strong pinning crossover}

\author{G.\ Blatter, V.B.\ Geshkenbein, and J.A.G.\ Koopmann}

\affiliation{Theoretische Physik, ETH-H\"onggerberg, CH-8093 Z\"urich,
Switzerland}

\date{\today}

\begin{abstract}
  Material defects in hard type II superconductors pin the
  flux lines and thus establish the dissipation-free current
  transport in the presence of a finite magnetic field.
  Depending on the density and pinning force of the defects
  and the vortex density, pinning is either weak-collective
  or strong. We analyze the weak- to strong pinning crossover
  of vortex matter in disordered superconductors and discuss
  the peak effect appearing naturally in this context.
\end{abstract}

\maketitle

\begin{figure}
   \includegraphics[scale=0.318]{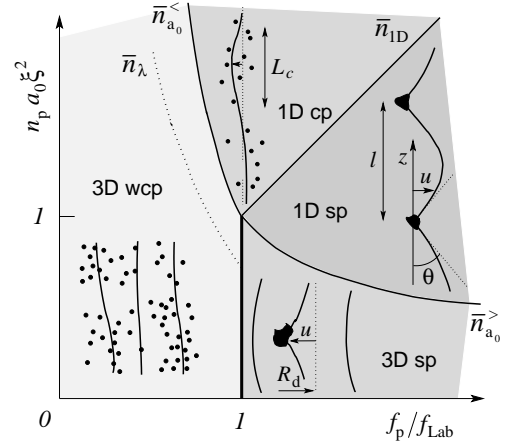}
   \caption[]{Pinning diagram delineating the 
   various pinning regimes involving collective-
   versus individual pinning and 1D-line- versus 3D-bulk
   pinning ($f_{\rm Lab}$ denotes the Labusch force):
   3D wcp -- bulk weak collective pinning, 1D cp --
   collective line pinning, 1D sp -- strong line pinning,
   3D sp -- bulk strong pinning. The insets illustrate
   the weak- and strong (plastic) distortions of the
   vortex system characterizing the pinning regimes. 
   Lines refer to crossovers.}
   \label{fig:w2s_1}
\end{figure}

Pinning of vortices by material defects is crucial in establishing
the defining property of a superconductor, its ability to
transport electrical current without dissipation. Collective
pinning theory \cite{LO_79}, describing the concerted action of
many weak pins on the vortex system, is playing a central role in
our understanding of this complex statistical mechanics problem
\cite{review}. On the other hand, first attempts describing flux
pinning go back to Labusch \cite{Lab_69}, who described the
interaction between vortices and strong pinning centers which
introduce large (plastic) deformations in the vortex system. In
this letter, we describe how these two theories relate to one
another; given the density $n_\mathrm{p}$ and force $f_\mathrm{p}$
of pinning centers, as well as the vortex density $n_\mathrm{v}=
1/a_0^2$, we identify the regimes where individual vortex lines
and the bulk vortex lattice are pinned by the collective action of
many weak pins or by the independent action of strong pins, see
Fig.\ \ref{fig:w2s_1}. We naturally recover the peak effect
\cite{pippard_69} described in the work of Larkin and Ovchinnikov 
\cite{LO_79} and establish its formal relation to the Landau 
theory of phase transitions.

In a type II superconductor, the field (${\bf B}$) induced vortices
subject to a current flow ${\bf j}$ experience the Lorentz force
density ${\bf F}_{\rm \scriptscriptstyle L} = {\bf j}\wedge{\bf B}/c$
and the resulting vortex motion leads to dissipation. The
superconducting response is resurrected through material
inhomogeneities pinning the vortices at energetically favorable
locations. The pinning force density ${\bf F}_\mathrm{pin}$ defines a
critical current density $j_c = c F_\mathrm{pin}/B$ below which the
current can flow free of dissipation. Usually, this critical current
density is considerably reduced with respect to the depairing current
density $j_0 \sim c H_c / 4 \pi \lambda \sim c \varepsilon_0/
\Phi_0\xi$; here, $H_c = \Phi_0/2\sqrt{2} \pi\lambda\xi$ is the
critical magnetic field, $\lambda$ and $\xi$ denote the penetration
depth and the coherence length $\xi$, respectively, $\Phi_0 = h c/2e$
is the flux unit, and $\varepsilon_0 = (\Phi_0/4\pi \lambda)^2$ is the
(line) energy scale. Below, we focus on the most generic situation of
isotropic superconductors and ignore effects due to thermal
fluctuations.

\begin{figure}
\includegraphics[scale=0.383]{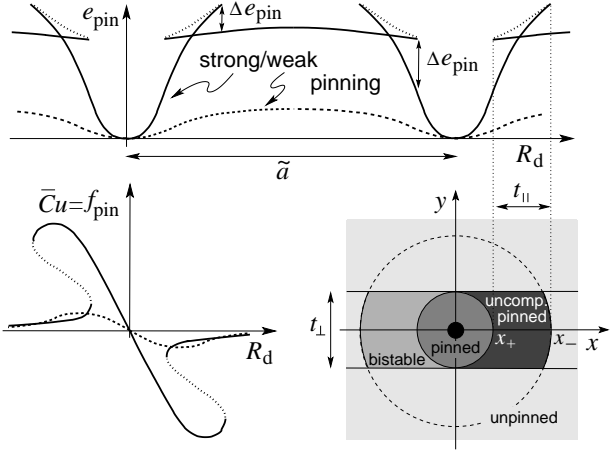}
\caption[]{Energy landscape $e_\mathrm{pin}$ and pinning force
$f_\mathrm{pin}$ versus displacement $R_\mathrm{d}$ of the vortex
lattice relative to the defect; for weak pinning these are
single-valued functions in $R_\mathrm{d}$ ({\itshape dashed
lines}), while strong pinning produces plastic deformations and
renders $e_\mathrm{pin}$, $f_\mathrm{pin}$ multi-valued ({\itshape
solid lines}; {\itshape dotted lines} indicate unstable branches).
Bottom right: Trapping geometry
(top view) for a circularly symmetric situation.}
\label{fig:w2s_2}
\end{figure}

When pinning is strong \cite{Lab_69,LO_79,LO_NSC_86}
defects act individually and the pinning force density
$F_\mathrm{p}$ is linear in the density $n_\mathrm{p}$ and
average pinning force $\langle f_\mathrm{pin} \rangle$ of
defects. The classic arguments characterizing strong pinning
go back to Labusch \cite{Lab_69}, see also \cite{LO_79,LO_NSC_86}:
A strong pinning defect induces plastic deformations in the
vortex lattice \cite{Brandt_86gen,OvIv_91,Schoenenberger_96})
and the energy landscape $e_\mathrm{pin}({\bf R})$ becomes
multi-valued in the vortex position ${\bf R}$, see 
Fig.\ \ref{fig:w2s_2}. The averaging over defect locations 
then has to account for the preparation of the system.
We concentrate on the critical current density and
thus search for the force against drag; the vortex 
position then is parametrized through the two-component
drag parameter ${\bf R}_\mathrm{d}$ fixing the position 
of the unperturbed lattice with respect to the defect. 
Dragging the system along the $x$-direction, we express 
the drag force $-\partial_x e_\mathrm{pin} (x,y)$ 
integrated along $x$ through the jump $\Delta 
e_\mathrm{pin}(y)>0$ in the pinning energy and 
average over `impact parameters'~$y$,
\[
   \langle f_\mathrm{pin} \rangle
   =-\int_0^{L_x}\!\!\!\! d x \int_0^{L_y} \!\!\!\!d y
   \frac{\partial_x e_\mathrm{pin}(x,y)}{L_x L_y}
   =-\int_0^{a_0}\!\!\! dy
   \frac{\Delta e_\mathrm{pin}(y)}{a_0 \tilde{a}(y)},
\]
where $\tilde{a}$ denotes the distance between periodic
jumps \cite{note}. For moderately strong pins with 
deformations not exceeding the lattice constant we 
have $\tilde{a} \approx a_0$ and assuming a maximal 
transverse trapping distance $t_\perp$ along the 
$y$-axis we obtain the mean pinning force
\begin{equation}
   \langle f_\mathrm{pin} \rangle
   \approx -\frac{t_\perp}{a_0^2} \Delta e_\mathrm{pin}(0)
   \approx -\frac{t_\perp t_\parallel}{a_0^2} f_\mathrm{p}
   \approx -\frac{S_\mathrm{trap}}{a_0^2} f_\mathrm{p},
   \label{fav}
\end{equation}
with the jump $\Delta e_\mathrm{pin}(0) \approx t_\parallel
f_\mathrm{p}$ expressed via the typical impurity force
$f_\mathrm{p}$ and the bistability range $t_\parallel$ of
$e_\mathrm{pin}(x,0)$; the product $t_\perp t_\parallel$
defines the trapping area $S_\mathrm{trap}$ associated with the
strong pin \cite{OvIv_91}. The low impurity concentration 
$n_\mathrm{p}$ implies non-interfering defects and we obtain 
a critical current density $j_c=-c n_\mathrm{p}\langle 
f_\mathrm{pin}\rangle/B$ linear in $n_\mathrm{p}$,
\begin{equation}
   j_c
   \approx ({c}/{B}) n_\mathrm{eff} f_\mathrm{p}
   \approx j_0 [n_\mathrm{p} \xi S_\mathrm{trap}]\,
   {f_\mathrm{p}}/{\varepsilon_0},
   \label{j_sp}
\end{equation}
with the effective impurity density $n_\mathrm{eff} = n_\mathrm{p}
(S_\mathrm{trap}/a_0^2)$.

In order to derive a quantitative criterion for the appearance of
strong pinning, we consider a single defect at the origin
with a pinning potential $e_\mathrm{p} ({\bf r})$. Such a defect
acts on the vortex system to produce a pinning energy density
$E_\mathrm{p}({\bf r},{\bf u}) = \sum_{\nu}e_\mathrm{p}
({\bf r}) \delta^2({\bf R}-{\bf R}_\nu -{\bf u}({\bf R}_\nu,z))$,
with vortices positioned at ${\bf R}_\nu+{\bf u}({\bf R}_\nu,z)$,
${\bf R}_\nu$ the equilibrium positions and ${\bf u}$ the
displacement field. The latter follows from the solution
of the implicit equation (${\bf r}_\nu=({\bf R}_\nu,z)$)
\begin{eqnarray}
   &&{u}_\alpha({\bf r}_\nu)
   = \!\! \int \! d^3 r^\prime\,
   G_{\alpha\beta}({\bf r}_\nu-{\bf r}^\prime)
   [-\partial_{u_\beta} {E}_\mathrm{p}]({\bf r}^\prime,{\bf u}^\prime)
   \nonumber \\
   &&\quad =
   \sum_{\nu^\prime} \int dz^\prime
   G_{\alpha\beta}({\bf r}_\nu-{\bf r}_{\nu}^\prime)
   {f}_{\mathrm{p},\beta}({\bf R}_{\nu}^\prime
   +{\bf u}({\bf r}_{\nu}^\prime),z^\prime)
   \nonumber \\
   &&\quad = G_{\alpha\beta}({\bf R}_\nu-{\bf R}_\mathrm{d},0)
   {f}_{\mathrm{p},\beta}({\bf R}_\mathrm{d}+
   {\bf u}({\bf R}_\mathrm{d},0),0),
   \label{u_int_f}
\end{eqnarray}
with $G_{\alpha\beta}(\bf r)$ the elastic Green's function and
${\bf f}_\mathrm{p} = -\nabla_u e_\mathrm{p} ({\bf u})$ the
pinning force of the defect. In the last equation we have assumed
a moderately strong pinning potential (pinning one vortex at most)
of range much smaller than the lattice constant $a_0$ and have
chosen ${\bf R}_\mathrm{d}$ as the distance to the vortex closest
to the defect \footnote{Effects of very strong pinning 
(overdrag into the next cell and multi-vortex pinning) 
are not considered here.}.
Evaluating (\ref{u_int_f}) for ${\bf r}_\nu =
({\bf R}_\mathrm{d},0)$, we arrive at the self-consistency equation
(note that $G_{\alpha\beta} ({\bf r}=0)$ is diagonal)
\begin{equation}
   {\bf u}({\bf R},0) \approx  \bar{C}^{-1} {\bf f}_\mathrm{p}
   ({\bf R}+{\bf u}({\bf R},0),0),
   \label{self}
\end{equation}
with the effective elastic constant $\bar{C}^{-1} = \int 
d^3 k/(2\pi)^3$ $G_{xx}({\bf k})$. 
For weak pinning the displacement ${\bf u}$ is small and the
solution ${\bf u}({\bf R},0) \approx {\bf f}_\mathrm{p} ({\bf
R})/\bar{C}$ is unique. Strong pinning, however, produces
multi-valued functions ${\bf u}({\bf R},0)$ and
${e}_\mathrm{pin}({\bf R})$, cf.\ Fig.\ \ref{fig:w2s_2}. The
solution of (\ref{self}) turns multi-valued as the displacement
collapses when $\partial_R u \rightarrow \infty$. Assuming a
defect symmetric in the plane, $e_\mathrm{p}({\bf R},z) =
e_\mathrm{p} (R,z)$, and dragging the lattice through the defect
center along the $x$-axis, we find $u' = f'_\mathrm{p}(x+u)$
$[\bar{C}-f'_\mathrm{p}(x+u)]^{-1}$ (note that $x>0$ implies $u<0$) and
arrive at the (Labusch) criterion \cite{Lab_69} in the form
\begin{equation}
   \partial_x f_\mathrm{p} = -\partial_x^2 e_\mathrm{p}
   = \bar{C};
   \label{sp_crit}
\end{equation}
hence, in order to produce strong pinning the (negative) curvature
of the pinning energy $e_\mathrm{p}$ has to overcompensate the
lattice elasticity (the Labusch criterion involves the maximal
negative curvature above the inflection point). Note that the
Labusch criterion tests an individual pinning center and
classifies it as a weak or strong one.

When pinning is weak, the elastic forces dominate over the pinning
forces and the defects compete; we then are faced with the problem
of the statistical summation of individual pinning forces. For
weak pins the average $\langle f_\mathrm{pin}\rangle$ vanishes and
pinning is due to fluctuations in the pinning force density: the
forces of the competing pins (with pinning force $f_\mathrm{p}$,
density $n_\mathrm{p}$, and extension $r_\mathrm{p} \sim \xi$) add
up randomly and produce the pinning energy
\begin{equation}
   \langle {\cal E}_\mathrm{pin}^2 (V) \rangle^{1/2}
   \approx \bigl[f_\mathrm{p}^2 n_\mathrm{p}
   (\xi/a_0)^2 V\bigr]^{1/2}
   \, \xi;
   \label{E2pin_g}
\end{equation}
only vortex cores are pinned by the disorder, hence the factor
$(\xi/a_0)^2$. Within weak collective pinning theory the sublinear
growth of $\langle {\cal E}_\mathrm{pin}^2 (V) \rangle^{1/2}$ with
volume turns linear when the displacement $u$ increases beyond the
scale $\xi$ of the pinning potential, thus defining the collective
pinning volume $V_c$. Each volume of size $V_c$ is pinned
independently with a pinning energy $U_c = \langle {\cal
E}_\mathrm{pin}^2 (V_c) \rangle^{1/2}$ and we obtain a proper
pinning force density
\begin{equation}
   F_\mathrm{pin}
   \sim {U_c}/{V_c \,r_\mathrm{p}}
   \sim \bigl({f_\mathrm{p}^2
   n_\mathrm{p}(\xi/a_0)^2}/{V_c}\bigr)^{1/2};
   \label{f_pin_g}
\end{equation}
balancing this pinning force density against the Lorentz force
density $j B /c$ we find a finite critical current density $j_c
\sim c F_\mathrm{pin}/B$. The remaining task is the determination
of the collective pinning volume $V_c$; its calculation is
complicated by the dispersion and anisotropy of the vortex
lattice, see below and Ref.\ \cite{review} for a detailed
discussion.

It is instructive to compare the weak- and strong pinning schemes
and their dependence on dimensionality, particularly in the limit
of a small defect density $n_\mathrm{p}$ (in the following, we
assume pinning sites characterized by their force $f_\mathrm{p}$
and extension $\xi$). An isolated vortex line (1D) is always
subject to strong pinning forces as the effective elastic
coefficient $\bar{C}$ vanishes due to the diverging integral.
At the same time, the deformation of the line due to
the pins is large and we cannot ignore their mutual competition.
Comparing the elastic energy $\varepsilon_0 \xi^2/L_c$ and the
pinning energy $U_c = (f_\mathrm{p}^2 n_\mathrm{p} L_c
\xi^2)^{1/2} \xi$, we find the collective pinning length $L_c \sim
(\varepsilon_0^2 /f^2_\mathrm{p} n_\mathrm{p})^{1/3}$ and a
critical current density
\begin{equation}
   j_c \sim j_0 \bigl({n_\mathrm{p}\xi^3
   f^2_\mathrm{p}}/{\varepsilon_0^2}\bigr)^{2/3}.
   \label{jc_1d_cp}
\end{equation}
This result is valid as long as many pins compete within the
volume $\xi^2 L_c$; the condition $n_\mathrm{p} \xi^2 L_c > 1$
defines the lower limit $\bar{n}_{\rm \scriptscriptstyle 1D} \sim
f_\mathrm{p} /\varepsilon_0 \xi^3 < n_\mathrm{p}$ where the
critical current density assumes the value $\bar{j_c} \sim j_0
(f_\mathrm{p}/\varepsilon_0)^2$.

For small densities $n_\mathrm{p} < \bar{n}_{\rm
\scriptscriptstyle 1D}$ the pins acts individually and we
determine $j_c$ from the force balance $(\Phi_0/c) j_c l u \sim
\Delta e_\mathrm{pin} \sim f_\mathrm{p} u$, with $u\sim
t_\parallel$ the displacement directed along the force. The
displacement $u$ and the length $l$ between two subsequent pins
fixing the vortex derives from an analysis of the pinned vortex
geometry, see Fig.\ \ref{fig:w2s_1} inset: integrating the force
equation $\varepsilon_0 {\partial_z^2 u} = f(z)$ (with $f(z)$ the
force per unit length acting on the line) over one pinning center,
we find the distortion angle $\theta =\partial_z u \sim u/l \sim
f_\mathrm{p}/\varepsilon_0$ \cite{alternative}. A vortex segment
of length $l$ deformed by the angle $\theta$ in the direction of
the driving force encounters $\theta l^2 \xi n_\mathrm{p}$ defects
(with the trapping length $t_\perp \sim \xi$). At the distance
$l$, this number is unity, hence $l \sim
\sqrt{\varepsilon_0/f_\mathrm{p} n_\mathrm{p} \xi}$ and we obtain
the critical current density
\begin{equation}
   j_c \sim j_0
   \bigl({n_\mathrm{p}
   \xi^3 f^3_\mathrm{p}}/{\varepsilon_0^3}\bigr)^{1/2}.
   \label{jc_1d_sp}
\end{equation}
At the crossover density $\bar{n}_{\rm\scriptscriptstyle 1D} \sim
f_\mathrm{p}/ \varepsilon_0 \xi^3$ the critical current density
matches up with the weak pinning result; also, the displacement $u
\sim l f_\mathrm{p}/\varepsilon_0$ is of order $\xi$ at the
crossover density $\bar{n}_{\rm \scriptscriptstyle 1D}$ and hence
matches the displacement field relevant in the collective pinning
scenario. Note that collective pinning (\ref{jc_1d_cp}) dominates
over the strong pinning (\ref{jc_1d_sp}) at \textit{large}
densities $n_\mathrm{p} > \bar{n}_{\rm\scriptscriptstyle 1D}$.

For the vortex lattice (3D bulk pinning; we assume $a_0<\lambda$)
the Labusch criterion (\ref{sp_crit}) offers a distinction between
weak and strong pinning centers; using the Green's function for
the vortex lattice (see, e.g., \cite{review}) we find $\bar{C} 
\sim \varepsilon_0/a_0$.
According to (\ref{sp_crit}) a pinning center changes from weak to
strong at $f_\mathrm{p} \sim f_\mathrm{Lab} \equiv \varepsilon_0
\xi/a_0$. We first review the weak pinning situation with
$f_\mathrm{p} < f_\mathrm{Lab}$ (where necessary, we encode
quantities in this regime with a superscript
`$^{\scriptscriptstyle <}$'). The determination of the anisotropic
collective pinning volume $V_c = R_c^2 L_c^\mathrm{b}$ has to
account for the dispersion in the tilt modulus at intermediate
scales $a_0 < R_c < \lambda$, see Ref.\ \cite{review}, and
produces the results
\begin{eqnarray}
   &&j_c \sim j_0 \,\frac{\xi^2}{a_0^2}
   \left[\frac{a_0}{L_c}\right]^\nu
   e^{-2c\left[L_c/a_0\right]^3} , \quad R_c < \lambda,
   \label{sbp}\\
   &&j_c \sim j_0 \,\frac{\xi^2}{\lambda^2}
   \left[\frac{a_0}{L_c}\right]^6,
   \quad R_c > \lambda;
   \label{lbp}
\end{eqnarray}
we have made use of the single vortex pinning parameter $L_c/a_0
\sim (f_\mathrm{Lab}^2/f_\mathrm{p}^2 a_0 \xi^2
n_\mathrm{p})^{1/3}$. The numericals $c$ and $\nu$ follow from a
2-loop renormalization group analysis \cite{wagner,book}. These
results are valid as long as many (competing) pins are active in
the volume $V_c$, $n_\mathrm{p}(\xi^2/ a_0^2) V_c > 1$. For
$f_\mathrm{p}< f_\mathrm{Lab}$ this condition is violated in the
large $n_\mathrm{p}$ limit. However, with increasing pinning
density $n_\mathrm{p}$, the collective pinning radius $R_c$
decreases, first below $\lambda$ at $\bar{n}_\lambda \sim
f_\mathrm{Lab}^2/f_\mathrm{p}^2 a_0\xi^2 \ln(\lambda/a_0)$
delineating the dispersive regime, and then below $a_0$ at
$\bar{n}_{a_0}^{\scriptscriptstyle <} \sim
f_\mathrm{Lab}^2/f_\mathrm{p}^2a_0\xi^2$ where the condition
$n_\mathrm{p}(\xi^2/ a_0^2) V_c > 1$ is still fulfilled. At the
crossover density $\bar{n}_{a_0}^{\scriptscriptstyle <}$ the 3D
weak collective pinning crosses over to the 1D weak collective
pinning result (\ref{jc_1d_cp}).

Turning to strong pinning $f_\mathrm{p} > f_\mathrm{Lab}$ (encoded
with a superscript `$^{\scriptscriptstyle >}$') we start at high
densities; as the Labusch criterion is not effective in 1D, the
system remains collectively pinned for $n_\mathrm{p}> \bar{n}_{\rm
\scriptscriptstyle 1D}$ and crosses over to 1D strong pinning as
$n_\mathrm{p}$ drops below $\bar{n}_{\rm \scriptscriptstyle 1D}$.
Decreasing $n_\mathrm{p}$ further, the pinning distance $l \sim
a_0\sqrt{(f_\mathrm{Lab}/ f_\mathrm{p})/n_\mathrm{p}a_0\xi^2}$
increases beyond $a_0$ as $n_\mathrm{p}$ decreases below
$\bar{n}_{a_0}^{\scriptscriptstyle >} \sim (f_\mathrm{Lab} /
f_\mathrm{p})/a_0\xi^2$ and the system enters the 3D strong
pinning regime, see Fig.\ \ref{fig:w2s_1}. The calculation of the
mean pinning force density $F_\mathrm{pin} \sim n_\mathrm{p}
\langle f_\mathrm{pin}\rangle$ proceeds along the lines 
discussed above and involves the trapping area $S_\mathrm{trap} 
\sim t_\perp t_\parallel$ with $t_\perp \sim \xi$ and
$t_\parallel \sim u \sim f_\mathrm{p}/\bar{C}$; we obtain the
force density $F_\mathrm{pin} \sim n_\mathrm{p} (\xi/a_0)
f^2_\mathrm{p} /\varepsilon_0$ and a critical current density
\begin{equation}
   j_c \sim j_0\,
   a_0 \xi^2 n_\mathrm{p}
   \frac{f^2_\mathrm{p}}{\varepsilon_0^2}
   \sim j_0\, \frac{\xi^2}{a_0^2}
   n_\mathrm{p} a_0 \xi^2 \,
   \frac{f^2_\mathrm{p}}{f_\mathrm{Lab}^2}.
   \label{jc_3d_sp}
\end{equation}
The bulk strong pinning result (\ref{jc_3d_sp}) smoothly
transforms into the 1D expression (\ref{jc_1d_sp}) at
$\bar{n}_{a_0}^{\scriptscriptstyle >}$ where $l \sim a_0$. On the
contrary, the strong pinning expression (\ref{jc_3d_sp}) does not
match up with the bulk weak collective pinning results (\ref{sbp})
and (\ref{lbp}) at $f_\mathrm{p} = f_\mathrm{Lab}$ (we concentrate
on low impurity densities with $n_\mathrm{p} a_0\xi^2 < 1$, cf.\
Fig.\ \ref{fig:w2s_1}). However, we have to keep in mind that our
rough derivation of the strong pinning result (\ref{jc_3d_sp})
breaks down on approaching the critical force $f_\mathrm{Lab}$.
Indeed, since the displacement field ${\bf u}({\bf r})$ turns
single valued below $f_\mathrm{Lab}$, strong pinning vanishes
altogether (with a power $[f_\mathrm{p}-f_\mathrm{Lab}]^2$, see
(\ref{jc_3d_sp_gen})) and pinning survives only in the form of
weak collective pinning due to fluctuations in the impurity
density. Within the approximative scheme adopted here the sharp
rise of the critical current density at $f_\mathrm{p} >
f_\mathrm{Lab}$ is encoded in a jump $j_c|_\mathrm{sp}/
j_c|_\mathrm{wcp} \sim (\lambda^2/a_0^2) /n_\mathrm{p} a_0 \xi^2 >
1$ for $n_\mathrm{p} < \bar{n}_\lambda$ ($\sim \exp[2c/
n_\mathrm{p} a_0 \xi^2]$ for $n_\mathrm{p} > \bar{n}_\lambda$).

The crossover from strong to weak pinning at the Labusch condition
(\ref{sp_crit}) can be analyzed within a Landau type expansion: We
define the free energy functional $e_\mathrm{pin}({\bf u},{\bf
R}_\mathrm{d}) = \bar{C} u^2/2 + e_\mathrm{p}({\bf R}_\mathrm{d}
+{\bf u})$ from which the self-consistency equation (\ref{self})
follows by variation. Note that the derivative $-\partial_x
e_\mathrm{pin} = f_{\mathrm{p},x}({\bf R}_\mathrm{d}+{\bf u})$ 
provides the force along $x$ acting on a vortex separated from 
the defect by ${\bf R}_\mathrm{d}$ and deformed by ${\bf u}$, 
c.f.\ Fig.\ 1; it is this force which has to be averaged 
over in the definition of $\langle f_\mathrm{pin}\rangle$.

We first concentrate on the trajectory ${\bf
R}_\mathrm{d} = (x,0)$ with ${\bf u} = (u,0)$. The curvature
$e_\mathrm{p}^{\prime \prime}(u)$ relevant in (\ref{sp_crit})
assumes a maximal negative value; we denote the corresponding
location and value by $u_\kappa$ and $-\kappa$, respectively.
Next, we expand the curvature around $u_\kappa$,
$e_\mathrm{p}^{\prime\prime}(u) \approx -\kappa + \alpha
(u-u_\kappa)^2/2$; integrating in $u$ and combining with 
the elastic term $\bar{C} u^2/2$ we arrive at the expansion
\begin{eqnarray}
   e_\mathrm{pin}[u,x]
   &\approx& \bar{C}\,u^2/2
   +\nu\,(x+u-u_\kappa)\label{L_expansion}\\
   &-&\kappa(x+u-u_\kappa)^2/2
   +\alpha(x+u-u_\kappa)^4/24.
   \nonumber
\end{eqnarray}
This pinning energy maps to the free energy $e_\mathrm{mag}
[\phi,h] = \tau \phi^2/2 +\alpha \phi^4 /24-h\phi$ of a 
one-component magnet in a magnetic field \cite{LL} 
if we define the `order parameter' $\phi = x+u-u_\kappa$, 
the `temperature difference' $\tau = \bar{C} -\kappa$, and 
the `magnetic field' $h = \bar{C}(x-u_\kappa -\nu/\bar{C})$.
The `high-temperature' phase $\tau>0$ describing the 
paramagnet corresponds to weak pinning, while the
two ferromagnetic phases at `low temperatures' $\tau < 0$ 
stand for the pinned ($\phi <0$) and unpinned ($\phi >0$) 
states; the transition between these states is discontinuous 
and the associated coexistence regime extends over the `field' 
domain $|h| < h^\ast = (2/3\bar{C})\sqrt{2/\alpha}|\tau|^{3/2}$.
At $h^\ast$ the trapping/detrapping of the vortex from the defect
produces jumps $\Delta\phi = 3\sqrt{2/\alpha}|\tau|^{1/2}$
in the `order parameter', leading to jumps $\Delta e = \Delta 
e_\mathrm{pin}/2=(9/2\alpha)\, \tau^2$ in the energy. 
 
For finite `impact parameters' $y$ we have to determine
the trapping distance $t_\perp$. Assuming rotational
symmetry, the bistable regime is bounded by a circle
of radius $R = x^\ast = u_\kappa +\nu/\bar{C}$ and 
hence $t_\perp \approx 2 x^\ast$ (note that at $\tau = 0$ 
we have $h^\ast = 0$ but the critical drag parameter 
$x^\ast$ does not vanish). The (uncompensated)
trapping area determining the average pinning force $\langle
f_\mathrm{pin}\rangle$ is shown in Fig.\ \ref{fig:w2s_2};
combining the above results for the jump in pinning energy 
and the transverse trapping distance we find the averaged 
pinning force (c.f.\ (\ref{fav}))
\begin{equation}
   \langle f_\mathrm{pin} \rangle 
   \approx 18(u_\kappa +\nu/\bar{C})[\bar{C}-\kappa]^2/\alpha a_0^2.
   \label{fpin_Lab}
\end{equation}
Defining the individual force of (equal) pinning centers 
via $f_\mathrm{p} = \max_u[\partial_u f_\mathrm{p}](u) \xi =
\kappa \xi$ (then $f_\mathrm{Lab} = \bar{C} \xi$)
we can translate (\ref{fpin_Lab}) into an expression
for the critical current density $j_c$ extending the strong 
pinning result (\ref{jc_3d_sp}) to the vicinity of the
Labusch point,
\begin{equation}
   j_c \sim j_0\,(\xi^2/a_0^2)
   n_\mathrm{p} a_0 \xi^2 \,
   [f_\mathrm{p}/f_\mathrm{Lab}-1]^2.
   \label{jc_3d_sp_gen}
\end{equation}
Comparing with the weak pinning result (\ref{lbp}),
we note a sharp rise in the critical current density $j_c$
once the strong pinning force overcomes the weak 
pinning result \cite{LO_79}. With the small parameter 
$\delta = (a_0/\lambda) \sqrt{n_\mathrm{p} a_0 \xi^2} < 1$, 
this crossover appears above but still close to the Labusch 
point as $f_\mathrm{Lab} \propto \bar{C}$ decreases 
below $f_\mathrm{p}/(1+\delta)$.

Another remarkable result is the interpretation of the collective
pinning scenario in terms of the strong pinning picture; indeed,
summing over competing pins within the collective pinning volume
$V_c$ produces the corresponding critical Labusch force.
Quantitatively, we compare the force gradient $f' \sim
[n_\mathrm{p}(\xi^2/a_0^2)V]^{1/2} f_\mathrm{p}/\xi$ accumulated
within the (anisotropic) volume $V = L R^2 = (\lambda/a_0)R^3$
with the elastic parameter $\bar{C}(R) =\varepsilon_0 \lambda
R/a_0^3$ for smooth distortions on the scale $R>\lambda$
(non-dispersive regime) and apply the Labusch criterion
(\ref{sp_crit}). We then find the scale $R_c = \lambda
f_\mathrm{Lab}^2/f_\mathrm{p}^2 n_\mathrm{p} a_0 \xi^2$, where the
accumulated pinning force overcompensates the elastic force; this
length agrees with the 3D collective pinning length in the
non-dispersive regime \cite{review}. The resulting bistable
solutions are the signature of the alternative pinning valleys
which the collective pinning volume can select beyond the scale
$R_c$.

The above discussion sheds light on the general concept of pinning.
Pinning is absent in the rigid limit. A finite but large elasticity
(with $f_\mathrm{Lab} > f_\mathrm{p}$) allows only for weak
deformations and individual pins cannot hold the lattice as the
averaging over individual pinning forces produces a null result. Hence,
pinning is due only to fluctuations in the pinning forces and thus
collective. Reducing the elasticity, strong pinning defects appear when
$f_\mathrm{Lab}$ drops below $f_\mathrm{p}$; they pin the lattice
individually and strong pinning, linear in the defect density
$n_\mathrm{p}$, outperforms collective pinning. The important role
played by the curvature $e_\mathrm{p}''<0$ in the pinning potential is
an interesting topic for  numerical studies. The crossover between weak
collective and strong pinning can be realized in  experiments:
increasing the magnetic field  towards its critical value $H_{c_2}$
leads to a marked softening of the elastic moduli. The reduction in the
elastic moduli entails a decrease of the Labusch force $f_\mathrm{Lab}$
and triggers the crossover from weak- to strong pinning, producing the
well known peak effect in the critical current density
\cite{LO_79,pippard_69}.

We acknowledge discussions with Anatoly Larkin and financial 
support from the Swiss National Foundation.


\begin{thebibliography}{99}
\bibitem{note} The integration along $x$ can be interpreted
   as an average of pinning forces or as the drag work over 
   the period $\tilde{a}$.
\bibitem{alternative} Alternatively, we find $\bar{C}$ using $G_{xx} =
1/\varepsilon_0 k_z^2$ and cut the integral on the scale
$l$, $\bar{C} \sim \varepsilon_0/l$; the relation $u \sim
f_\mathrm{p} /\bar{C}$ provides the angle $\theta \sim u/l \sim
f_\mathrm{p}/\varepsilon_0$.
\end{thebibliography}
\end{document}